\documentstyle[preprint,prb,aps]{revtex}
\begin{document}
\draft
\title{Randomly inhomogeneous Luttinger liquid: Fluctuations of the
 tunnel conductance.}
\author{A. Gramada \cite{pos} and M. E. Raikh}
\address{Department of Physics, University of Utah, Salt Lake City, 
Utah  84112}

\maketitle

\begin{abstract}

The Luttinger liquid in which the  
 concentration of electrons  varies randomly with coordinate is considered.
We study the fluctuations of the tunnel conductance, caused by the
randomness in the concentration. If the concentration changes slowly
on the scale of the Fermi wavelength, its prime role reduces to the
scattering  of the plasmon waves, propagating along the system.
As a result of such a scattering, plasmons get localized.
 We show that the localization length, $l_{\omega}$, of a
plasmon with frequency $\omega$ is inverse proportional to the square of
the interaction strength and  changes with frequency
  as $l_{\omega}\propto \omega^{-2}$.
If the  relative variation of the  concentration is small,
 the randomness--induced correction to the tunnel conductance,
$\delta G(V)$, where $V$ is the applied  bias,  can
be expressed through the spectral characteristics of the 
localized plasmons.  The magnitude of the correction, $(\overline{ 
\delta G^2})^{1/2}/ G$,
increases with $V$ as $\sqrt V$.
The typical period of the fluctuations 
 in $\delta G(V)$ is of the order of  $V$.
At a fixed $V$, the correlator of $\delta G$ at different points 
of the liquid falls off 
with distance as a power law and oscillates with the period which 
is one half of the wavelength of a plasmon with frequency $\omega = eV/\hbar$. 
\end{abstract}
\section{Introduction}
\label{sec:I}

It is well--known that the density of states
 in a pure one--dimensional
interacting electron gas (Luttinger liquid)
vanishes as a power law in the
vicinity of the Fermi level \cite{Lut}: $\nu(\omega)\propto
 \omega^{\kappa}$, where the exponent  is determined by 
the interaction strength.
Such a behavior should reveal itself
in the dependence of the differential tunnel conductance on the
applied bias: $G(V)= dI/dV \propto \nu (eV)$.
It is obvious that the presence of a disorder would
perturb the local value of the density of states and, thus,
 cause some random correction, $\delta G(V)$, to the conductance. 
Then the relevant questions  are: 

i) What is the typical magnitude
 of $\delta G$ for
a given realization of the disorder? 

ii) How the values of $\delta G$ at 
different voltages are correlated? 

iii) How the correlation between the values of 
$\delta G$ at the same voltage, but at different 
points of the liquid, falls off with increasing separation between the
points?  

These questions are addressed in the present paper.  We will 
consider the case of a smooth disorder. Namely, we will assume that the
correlation radius is much larger than the   
Fermi wavelength. This assumption  simplifies the problem drastically,
since it permits one   to neglect the backward scattering 
of electrons \cite{O&F}
 and, hence,
to  view the disorder  as a  random 
variation of the electron concentration with coordinate.
The key to understanding the role of the disorder is 
provided by  the
bosonization procedure \cite{Mat,Hal}, which allows to describe the
  low--energy excitations of the system
 in terms  of bosonic excitations (plasmons), propagating along the  
liquid. Since the velocity of a plasmon depends on the concentration
of electrons,  the spatial variation 
of the concentration  would give rise to the backscattering
of plasmons. In other words, in the presense
of  a disorder, a plasmon with frequency $\omega$
acquires a finite mean free path $l_{\omega}$. It is important to note
that a smooth disorder, for which the backscattering of electrons 
at the Fermi level is suppressed, might, ultimately, be  quite efficient in 
backscattering of plasmons with  wavelengths of  the order of the 
correlation radius.

For non--interacting
electrons in one dimension it is established that even
a weak disorder leads to the  localization
 of all  eigenstates \cite{Mot,Ber}. Then the mean free path acquires
the meaning of the localization radius. The same conclusion applies,
certainly, to  plasmons. The difference is, however,  that an
electron becomes more and more localized as its energy decreases, whereas
for plasmons the situation is the opposite: the lower is the frequency,
the weaker is the localization. Since the dispersion law of a plasmon
is linear: $\omega=sk$, where $s$ is the sound velocity, the density of
the plasmon states is frequency--independent. The reason for suppression
of the localization for the low--frequency plasmons is that the matrix 
element of the baskscattering vanishes in the limit of long wavelengths.
In particular, we show that for small $\omega$ the frequency dependence
of the mean free path is $l_{\omega}\propto \omega^{-2}$ and, 
correspondingly, the product $kl_{\omega}$, which determines
 the localization strength, increases as $1/\omega$ when $\omega$ goes
to zero. We also show that $l_{\omega}$ turns to infinity when the
interactions are switched off, thus revealing that the backscattering
of plasmons is possible only due to interactions. 

 If the relative
variation of the concentration is small, the disorder--induced 
correction to $\nu(\omega)$ can be expressed  in terms of the 
spectral characteristics  of the localized plasmons (such as
local density of states). On the other hand, these characteristics 
were  the subject of detailed studies  in  application to the localized 
electrons \cite{BG,GDP,NPF}. By utilizing the approach developed in
Refs.\onlinecite{BG,GDP}, we calculate the variance $(\overline{\delta
 G^2 })^{1/2}$, and  the two--point correlator
$\overline{\delta G(x)\delta G(0)}$. We  find that the 
ratio $(\overline{\delta G^2})^{1/2}/G$
increases with voltage as $\sqrt{V}$ and that the characteristic  
``period'' of change of $\delta G(V)$ with voltage is of the order of $V$.
We also find that the two--point correlator falls off with $x$ as a
power law and oscillates with the period $\delta x = \pi s\hbar/eV$, 
 which is one half of the wavelength of a plasmon with 
frequency $\omega = eV/\hbar$.

The paper is organized as follows.
In the next section the formula for the mean free path of a plasmon 
is derived.
In Section \ref{CTC} we calculate the correlator of  
fluctuations of the tunnel conductance. 
Section \ref{Con} concludes the paper.

\section{Mean free path of a plasmon}
\label{MFP}
The Hamiltonian of a Luttinger liquid with concentration of electrons,
$n(x)$, being a function of coordinate has the form \cite{Na,GR} 
\begin{equation}
\label{H=}
\hat{\cal H}= \int_{0}^{\infty}dx\biggl[\frac{\hat{p}^2(x)}{2mn(x)} 
+ \frac{1}{2}
\biggl(V_0 + \frac{\pi^2\hbar^2}{m}n(x)\biggr)
\biggl(\frac{d(n\hat{u})}{dx}\biggr)^2\biggr] ,
\end{equation} 
where $u$ and $\hat{p}(x)$ are, correspondingly, 
the displacement  
and conjugate momentum
($[\hat{u}(x),\hat{p} (x')]=i\hbar \delta(x-x')$);
$V_0$ is the effective  interaction strength: $V_0= \int dx V(x)$.
The Hamiltonian can be reduced to a system of harmonic oscillators 
\begin{equation}
\label{HH}
\hat{\cal H}= \sum_{\mu}\biggl[\frac{\hat{{\cal P}}_{\mu}^2}{2m}+
\frac{m\Omega_{\mu}^2}{2}\hat{Q}_{\mu}^2\biggr] ,                             
\end{equation}
by means of the following transformation
\begin{equation}
\label{p=}
\hat{u}(x)= \sum_{\mu}\frac{1}{\sqrt{n(x)}}\Phi_{\mu}(x)\hat{Q}_{\mu} \; ,
\hspace{1cm} 
\hat{p}(x)= \sum_{\mu}\sqrt{n(x)}\Phi_{\mu}(x)\hat{{\cal P}}_{\mu}.
\end{equation}
Here $\Phi_{\mu}$ are the eigenfunctions of the operator $\hat{D}$
which is defined as 
\begin{equation}
\label{D}
\hat{D} \Phi_{\mu}= -\sqrt{n(x)}\frac{d}{dx}\biggl[\biggl(\frac{V_0}{m}+ 
\left(\frac{\pi\hbar}{m} \right)^2 \!
n(x)\biggr)\frac{d}{dx} \left(\sqrt{n(x)}\Phi_{\mu} \right)\biggr]=
\Omega_{\mu}^2 \Phi_{\mu}.
\end{equation}
The eigenvalues of $\hat{D}$ determine the frequencies,
$\Omega_{\mu}$, of the oscillators. 
If the concentration is constant ($n(x)=n_0$), the solutions of (\ref{D})
are the plane waves
\begin{equation}
\label{PW}
\Phi_{\mu}^0 = \frac{e^{i k x}}{\sqrt{L}} \; ,
\end{equation}
with  a linear spectrum 
$\Omega_{\mu}=\Omega_{\mu}^0= s k_{\mu}$
($L$ is  the normalization length).
Substituting (\ref{PW}) into (\ref{D}), we get the standard expression 
for the sound velocity \cite{Mah}
\begin{equation}
\label{SV}
s= v_F \sqrt{1+\frac{V_0}{\pi \hbar v_F}} ,
\end{equation}
where $v_F =\pi \hbar n_0/m$ is the Fermi velocity.

Assuming that the relative variation of the concentration is small,
$|n(x)-n_0| \ll n_0$, the expression for the mean free time, 
$\tau_{\omega}$, for a plasmon with frequency $\omega$  is given
by  the golden rule
\begin{equation}
\label{GR}
\frac{1}{\tau_{\omega}}= \frac{\pi}{\omega}
\sum_{\nu} \overline{\left| \langle 
\Phi_{\mu}^{(0)} \left| \hat{D} \right| \Phi_{\nu}^{(0)} \rangle  \right|^2}
\delta(\Omega_{\nu}^{(0)^2}-\omega^2) \, ,
\end{equation}
where $\Phi_{\mu}^{(0)}$ is the plane wave with $k_{\mu}=\omega /s$ and 
$\overline{(\cdots)}$ stands for the averaging over the fluctuations of $n(x)$.
The factor $1/\omega$ in (\ref{GR}) appears since we define 
the mean free time as $1/\tau_{\omega}= \hbox{Im} \Omega_{\mu}=
\hbox{Im} \Omega_{\mu}^2/2 \omega $.
 Keeping only the first order terms in the difference $n(x)-n_0$, 
we get the following expression for the matrix element
\begin{eqnarray}
\label{ME}
\langle \Phi_{\mu}^{(0)} \left| \hat{D} \right| \Phi_{\nu}^{(0)} \rangle &=&
\frac{V_0}{2m} \int_{-\infty}^{+\infty}dx \, (n-n_0) 
\left( \Phi_{\mu}^{(0)^*} 
\frac{d^2 \Phi_{\nu}^{(0)}}{dx^2}+ 
\Phi_{\nu}^{(0)} \frac{d^2 \Phi_{\mu}^{(0)^*}}{dx^2} \right)  
\nonumber \\ 
 +\left( \frac{\pi \hbar}{m} \right)^2 &n_0& 
\int_{-\infty}^{+\infty}dx \, (n-n_0) \left[
\frac{1}{2} \Phi_{\mu}^{(0)^*}\frac{d^2 \Phi_{\nu}^{(0)}}{dx^2}+
\frac{1}{2} \Phi_{\nu}^{(0)} \frac{d^2 \Phi_{\mu}^{(0)^*}}{dx^2} -
\frac{d \Phi_{\mu}^{(0)^*}}{dx} \frac{d \Phi_{\nu}^{(0)}}{dx} \right] .
\end{eqnarray}
The energy conservation, insured by the $\delta$-function in (\ref{GR}),
requires that $k_{\nu}=-\omega / s$ (backscattering).
Then we have $\Phi_{\mu}^{(0)^*}=\Phi_{\nu}^{(0)}$ and, 
as it can be easily seen, the second term in the matrix element 
vanishes identically. As a result, the matrix element takes the form 
\begin{equation}
\label{FME}
\langle \Phi_{\mu}^{(0)} \left| \hat{D} \right| \Phi_{\nu}^{(0)} \rangle=
\frac{V_0}{mL} \left(\frac{\omega}{s} \right)^2 
\int_{-\infty}^{+\infty} dx \, (n-n_0) e^{2i \omega x/s} .
\end{equation} 
We see that the matrix element for backscattering is proportional
to the interaction strength, which reflects the fact that this process
is possible only due to interactions.
The mean free path, $l_{\omega}$, defined as $l_{\omega}=s \tau_{\omega}$, 
can be found after substituting (\ref{FME}) into (\ref{GR})
\begin{equation}
\label{MF}
\frac{1}{l_{\omega}}=\left( \frac{s^2-v_F^2}{s^2} \right)^2 
\frac{\overline{(\delta n)^2}}{2n_0^2} \left(\frac{\omega}{s} \right)^2
\int_{-\infty}^{+\infty}dx 
K \left(\frac{|x|}{R_c} \right) e^{2i \omega x/s}
\end{equation}
where we have introduced the correlator of the fluctuations of $n(x)$
\begin{equation}
\label{DC}
\overline{ (n(x)-n_0) (n(x^{\prime})-n_0)} =\overline{(\delta n)^2}
K \left( \frac{|x-x^{\prime}|}{R_c} \right).
\end{equation}
Here $R_c$ is the correlation radius and $\overline{(\delta n)^2}$ 
is the mean square fluctuation of $n(x)$, so that $K(0)=1$.

It is instructive to rewrite (\ref{MF}) for 
the product $kl_{\omega}=\omega l_{\omega}/s$,
which measures the effective scattering strength. 
Assuming the gaussian form of the 
correlator, $K(z)=e^{-z^2}$, we obtain
\begin{equation}
\label{MFG}
(kl_{\omega})^{-1}=\sqrt{\pi}\left( \frac{s^2-v_F^2}{s^2} \right)^2 
\frac{\overline{(\delta n)^2}}{2n_0^2}
\left(\frac{\omega R_c}{s} \, e^{-\omega^2 R_c^2/s^2} \right).
\end{equation}
Note that the last factor in (\ref{MFG}) 
is a function of the argument $\omega R_c/s$
and this function  does not exceed  unity. 
Thus, the product $kl_{\omega}$ is large for any $\omega$.
Obviously, the backscattering is ineffective  when $\omega R_c/s >1$,
i.e. when the correlation radius $R_c$ exceeds the wave length 
of the plasmon.
In the opposite limit, $\omega R_c/s <1 $, the mean free path behaves 
as $l_{\omega} \sim \omega^{-2}$. As it was mentioned in the
Introduction, the real meaning of $l_{\omega}$ is the localization 
length of a plasmon with frequency $\omega$. The fact that $kl_{\omega}>1$
allows one to apply to localized plasmons, the description developed 
earlier for localized electrons \cite{Ber,BG} just in this limit.
This is done in the next section.

\section{Correlator of the fluctuations of the tunnel conductance}
\label{CTC}
Once the Hamiltonian is diagonalized, the derivation of the formula 
for the density of states becomes  standard \cite{Mah}.
The operator $\Psi^{\dagger}(x_0)$, creating an electron at point $x_0$,
can be presented  in the form \cite{Mah,Hal}
\begin{equation}
\label{Bos}
\Psi^+(x_0)=exp \left\{-\frac{i}{\hbar} \sum \limits_{\mu} 
\left( \alpha_{\mu}(x_0) \hat{\mathcal{P}}_{\mu}+ 
\beta_{\mu}(x_0) \hat{Q}_{\mu} \right) \right\} ,
\end{equation}
where the coefficients $\alpha_{\mu}$ and $\beta_{\mu}$ are defined as
\begin{equation}
\label{AB}
\alpha_{\mu}(x_0)= \int_{x_0}^{+\infty} dx \frac{\Phi_{\mu}}
{\sqrt{n(x)}} \; , \hspace{0.5cm}
\beta_{\mu}(x_0)= \pi \sqrt{n(x_0)} \; \Phi_{\mu}(x_0) .
\end{equation}
The calculation of the Green function $\langle \Psi(t) 
\Psi^{\dagger}(0)\rangle $ does not differ from that for a pure 
Luttinger liquid and leads to the following expression for the density 
of states 
\begin{equation}
\label{DOS}
\nu(\omega)=\frac{1}{\pi}{\mbox Re}\int_0^{\infty}dt e^{i\omega t}
\langle \Psi(t)\Psi^{\dag}(0) \rangle = 
\frac{1}{\pi}{\mbox Re}\int_0^{\infty}dt e^{i\omega t}e^{-W(t)},
\end{equation} 
where $W$ is the sum over eigenmodes
\begin{equation}
\label{W=}
W(t)=\sum_{\mu} \left(\frac{m|\alpha_{\mu}|^2\Omega_{\mu}}{2\hbar}+
\frac{\hbar |\beta_{\mu}|^2}{2m \Omega_{\mu}} \right)
\biggl(1-e^{-i\Omega_{\mu}t}\biggr).
\end{equation} 
If $n(x)$ is constant so that the eigenfunctions are given by (\ref{PW}),
Eq.(\ref{DOS}) reproduces the known result \cite{Mah} for the density 
of states: $\nu_0(\omega) \propto \omega^{\kappa}$, with 
\begin{equation}
\label{kappa}
\kappa=\frac{1}{2} \left(\frac{s}{v_F}+\frac{v_F}{s} \right)-1=
\frac{1+\frac{V_0}{2 \pi \hbar v_F}}{\sqrt{1+\frac{V_0}{\pi \hbar v_F}}}-1
\end{equation} 
Indeed, for plane waves we have $|\alpha_{\mu}|^2=1/k_{\mu}^2Ln_0, 
|\beta_{\mu}|^2=\pi^2 n_0/L$ and we get the following expression 
for the function $W$ 
\begin{equation}
\label{WW}
W_0(t)=\frac{1}{2} \left(\frac{s}{v_F}+\frac{v_F}{s} \right) 
\int_0^{\infty}\frac{dk}{k}
(1-e^{-iskt})e^{-r_0k}=
\frac{1}{2} \left(\frac{s}{v_F}+\frac{v_F}{s} \right)
\ln\biggl(\frac{r_0 + ist}{r_0}\biggr),
\end{equation}
where $r_0$ is the cutoff parameter.
Substituting this expression into (\ref{DOS}) leads to (\ref{kappa}).

To find the disorder-induced correction to $\nu_0(\omega)$, we treat 
the difference $W(t)-W_0(t)$ as a perturbation and expand the exponent 
in (\ref{DOS}) to the first order. This gives 
\begin{equation}
\label{dDOS}
\delta\nu = -\frac{1}{\pi}{\mbox Re}\int_0^{\infty}dt e^{i\omega t}
\biggl(\frac{r_0}{r_0+ist}\biggr)^{\kappa+1}\biggl[W(t)-W_0(t)\biggr].
\end{equation}
To calculate the integral over $t$, it is convenient to present the 
denominator in (\ref{dDOS}) in the form
\begin{equation}
\label{Ga}
\frac{1}{(r_0 + ist)^{\kappa+1}}=\frac{1}{\Gamma(\kappa+1)}
\int_0^{\infty}dzz^{\kappa}e^{-z(r_0 + ist)}.
\end{equation}
After substituting (\ref{Ga}) into (\ref{dDOS}), both integrations,
over $z$ and $t$ can be easily carried out, and one obtains
\begin{eqnarray}
\label{main}
\delta\nu(x_0,\omega) = \frac{1}{\Gamma(\kappa+1)}
\biggl(\frac{r_0}{s}\biggr)^{\kappa+1}
\left\{ \sum \limits_{\mu} \left[\frac{m|\alpha_{\mu}|^2\Omega_{\mu}}{2 \hbar}
+ \frac{\hbar |\beta_{\mu}|^2}{2 m \Omega_{\mu}} \right]
(\omega-\Omega_{\mu})^{\kappa} \right.  \nonumber \\ 
\left.- (\kappa+1) \int_0^{\omega/s}\frac{dq}{q}(\omega-sq)^{\kappa} \right\} .
\end{eqnarray} 
Note that the constant $\alpha_{\nu}$ can be expressed through the 
derivative   $\frac{d\Phi_{\nu}}{dx}\bigg|_{x=x_0}$ by integrating  
the equation (\ref{D})  from $x_0$ to $\infty$.
Then, one gets
\begin{equation}
\label{al=}
\alpha_{\mu}= -\frac{s^2}{\Omega_{\mu}^2} \frac{1}{\sqrt{n_0}}
\left(\frac{d \Phi_{\mu}}{dx} \right)\Bigg |_{x=x_0} .
\end{equation}

To study the correlation properties of $\delta \nu $, we introduce 
the following local densities
\begin{eqnarray}
\label{Dens}
\rho_1(\varepsilon,x_0) &=& \sum \limits_{\mu} \left|
\Phi_{\mu}(x_0) \right|^2 \delta(\varepsilon - \Omega_{\mu}) , \\
\rho_2(\varepsilon,x_0) &=& s^2 \sum \limits_{\mu} 
\frac{1}{\Omega_{\mu}^2}\left| 
\frac{d \Phi_{\mu}(x_0)}{dx} \right|^2 
\delta(\varepsilon - \Omega_{\mu}) .
\end{eqnarray} 
The average values of these densities are equal
\begin{equation}
\label{AvDens}
\overline{ \rho_1}=
\overline{\rho_2} = \frac{1}{\pi s} . 
\end{equation}
Then the correction $\delta \nu$ can be rewritten in terms of 
the fluctuations $\delta \rho_1=\rho_1-1/\pi s$ and 
$\delta \rho_2=\rho_2-1/\pi s$
\begin{equation}
\label{FdDOS}
\delta\nu(x_0,\omega) = \frac{1}{\Gamma(\kappa+1)}
\biggl(\frac{r_0}{s}\biggr)^{\kappa+1} \frac{\pi s}{2}
\int_0^{\omega} \frac{d\omega_1}{\omega_1}
(\omega - \omega_1)^{\kappa}
\left[ \frac{v_F}{s} \delta \rho_1(\omega_1,x_0)+
\frac{s}{v_F} \delta \rho_2(\omega_1,x_0) \right] .
\end{equation}
Correspondingly, the correlator of $\delta \nu$ at different points 
is expressed through the correlators of the fluctuations  
$\delta \rho_1,\delta \rho_2$
\begin{eqnarray}
\label{NUCOR}
\overline{\delta\nu(x_1,\omega) \delta\nu(x_2,\omega)} &=& 
\left[\frac{1}{\Gamma(\kappa+1)}
\left(\frac{r_0}{s}\right)^{\kappa+1} \frac{\pi s}{2} \right]^2
\times \nonumber \\
& &\int_0^{\omega} \frac{d\omega_1}{\omega_1}
\int_0^{\omega} \frac{d\omega_2}{\omega_2} 
(\omega - \omega_1)^{\kappa}
(\omega - \omega_2)^{\kappa} 
\left(\frac{v_F^2}{s^2} F_1+
\frac{s^2}{v_F^2} F_2 +
2 F_{12} \right) \; ,
\end{eqnarray}
where the correlators $F_1,F_2$ and $F_{12}$ are defined as
\begin{eqnarray}
\label{COR}
F_1&=&\overline{\delta\rho_1(x_1,\omega_1) \delta\rho_1(x_2,\omega_2)} \; , \\
F_2&=&\overline{\delta\rho_2(x_1,\omega_1) \delta\rho_2(x_2,\omega_2)} \; , \\
F_{12}&=&\overline{\delta\rho_1(x_1,\omega_1) \delta\rho_2(x_2,\omega_2)} \; .
\end{eqnarray}

The correlator $F_1$ is, in fact, the correlator of the fluctuations 
of the local density. In application to localized electrons, 
it was studied in Refs. \onlinecite{NPF,GDP} using, correspondingly,
the Berezinski\v{i} technique \cite{Ber} and the technique 
developed by Berezinski\v{i} and Gor'kov \cite{BG}.
In the limit $\omega_1 \tau_{\omega_1} \gg 1, 
\omega_2 \tau_{\omega_2} \gg 1$, the correlator is nonzero 
only if the difference $\omega_1-\omega_2$ is small enough:
$|\omega_1-\omega_2| \sim 1/\tau_{\omega_1}$.
For distances $z=|x_1-x_2| \ll l_{\omega_1},l_{\omega_2}$, 
the expression for $F_1$, obtained in Ref. \onlinecite{NPF}, reads
\begin{equation}
\label{F1}
F_1= \frac{1}{3 \pi^2 s^2} \left[ \frac{\pi}{2 \tau_{\omega_1}}
\delta (\varepsilon) \left( 3-2 \sin^2\left(\frac{\omega_1 z}{s}\right) \right)
+2 \sin^2 \left(\frac{\omega_1 z}{s} \right) (C(\varepsilon)-1) \right] \; ,
\end{equation}
where $\varepsilon =\omega_1- \omega_2 $ 
is the difference between the two frequencies and the function $C(\omega)$ 
is given by 
\begin{equation}
\label{C}
C(\varepsilon)=(2 \varepsilon \tau_{\omega_1})^2 
\int_0^{\infty} dy \frac{\cos(2 \varepsilon \tau_{\omega_1}y)}{y+1}
= (2 \varepsilon \tau_{\omega_1})^2 \int_0^{\infty} dq 
e^{-q} \frac{q}{q^2 + 4 \varepsilon^2 \tau_{\omega_1}^2} \; .
\end{equation} 
As it was noted in Refs. \onlinecite{GDP,NPF,A.P}, at $z=0$ and 
$ \varepsilon\ne 0$, we have $F_1=0$, i.e. the correlation is absent.

It is easy to establish that the correlator $F_2$ is equal to $F_1$.
Concerning the correlator $F_{12}$, we did not find the expression for 
this correlator in the literature. So, we have calculated  it using the 
Berezinski\v{i}-Gor'kov technique and obtained the following expression
\begin{equation}
\label{F12}
F_{12}= \frac{1}{3 \pi^2 s^2} \left[ \frac{\pi}{2 \tau_{\omega_1}}
\delta (\varepsilon) \left( 1+2 \sin^2\left(\frac{\omega_1 z}{s}\right) \right)
+2 \cos^2 \left(\frac{\omega_1 z}{s} \right) (C(\varepsilon)-1) \right] \; .
\end{equation} 
It is seen that in contrast to $F_1$, the correlator $F_{12}$ is non-zero 
at $z=0$ and $\varepsilon \ne 0$ \cite{Note}. For finite $\varepsilon$,
both correlators are proportional to $(C(\varepsilon)-1)$ and decay
with increasing $\varepsilon$ as $(\varepsilon \tau_{\omega_1})^{-2}$.

To calculate the double integral in (\ref{NUCOR}), we make use of the 
fact that the correlators $F_1, F_2$ and $F_{12}$ are sharp functions 
of $\varepsilon$, i.e. the major contribution to the integral 
comes from the domain $|\omega_1-\omega_2| \sim 1/\tau_{\omega_1}
\ll \omega_1,\omega_2 $. 
This allows to put $\omega_2=\omega_1$ in all other factors 
and to extend  the integration over $\varepsilon=
\omega_1-\omega_2 $ to $(-\infty,\infty)$. 
Noting that the integral of $(1-C(\varepsilon))$ is equal to 
$\pi/\tau_{\omega_1}$, we obtain
\begin{equation}
\label{Int_e}
\int_{-\infty}^{\infty} d \varepsilon \left(\frac{v_F^2}{s^2} F_1+
\frac{s^2}{v_F^2} F_2 +2 F_{12} \right)=
\frac{1}{2 \pi s^2 \tau_{\omega_1}}\left(\frac{s}{v_F}-\frac{v_F}{s}\right)^2
\cos \left(\frac{2 \omega_1 z}{s} \right) \; .
\end{equation} 
Note that the coefficients in front of $F_1, F_2$ and $F_{12}$ in
(\ref{NUCOR}) have combined into the factor $(s/v_F- v_F/s)^2$
which is proportional to $V_0^2$ at small $V_0$. Using (\ref{Int_e}),
the expression for the correlator $\overline{\delta \nu(x_1,\omega)
\delta \nu(x_2,\omega)}$ takes the form
\begin{equation}
\label{nNUCOR}
\overline{\delta \nu(x_1,\omega) \delta \nu(x_2,\omega)}=
\frac{\pi}{8}
\left(\frac{s}{v_F}-\frac{v_F}{s}\right)^2
\left[\frac{1}{\Gamma(\kappa+1)}
\left(\frac{r_0}{s}\right)^{\kappa+1} \right]^2
\int_0^{\omega} \frac{d \omega_1}{\omega_1^2 \tau_{\omega_1}}
(\omega -\omega_1)^{2 \kappa} \cos \left(\frac{2 \omega_1 z}{s} \right) \; .
\end{equation} 
Apparently, the factor $1/\omega_1^2$ in (\ref{nNUCOR}) diverges 
at small $\omega_1$. However, this divergency is compensated by 
the frequency dependence of 
the mean free  time, which at small $\omega_1$ behaves as $1/\omega_1^2$.
Substituting $\tau_{\omega_1}$ in (\ref{nNUCOR}) and introducing the 
new variable $w =\omega_1/\omega$, we get the final result 
\begin{equation}
\label{nnNUCOR}
\overline{\delta \nu(x_1,\omega) \delta \nu(x_2,\omega)}=
\frac{\pi^{3/2}}{16 (2 \kappa + 1)}
\left[\frac{1}{\Gamma(\kappa+1)}
\left(\frac{r_0}{s}\right)^{\kappa+1} \right]^2
\left(1-\frac{v_F^2}{s^2}\right)^4
\frac{\overline{(\delta n)^2}}{n_0^2}
\frac{s R_c \omega^{2\kappa+1}}{v_F^2} 
R(\tilde{z}) \; ,
\end{equation} 
where the function $R$ is defined as 
\begin{equation}
\label{R}
R(\tilde{z})= (2\kappa+1) 
\int_0^1 dw (1-w)^{2\kappa} e^{-w^2 (\frac{\omega R_c}{s})^2}
\cos(w \tilde{z}) \; .
\end{equation} 
Here, $\tilde{z}$ is the dimensionless distance 
\begin{equation}
\label{z}
\tilde{z}=\frac{2\omega}{s} (x_2-x_1) \; .
\end{equation} 
The function $R$ is defined in such a way that, in the only interesting 
limit $\omega R_c/s \ll 1$, it turns to $1$ at $\tilde{z}=0$. 
This function determines the coordinate dependence of the correlator 
(\ref{NUCOR}) and, consequently, the coordinate dependence 
of the correlator  of the fluctuations of the tunnel conductance 
\begin{equation}
\label{TCF}
\frac{\overline{\delta G(x) \delta G(0)}}{\overline{(\delta G(0))^2}}=
R\left(\frac{2 e V}{s \hbar} x \right) \; .
\end{equation} 
The function $R$ is ploted in Fig.1 . With increasing distance, 
it falls off and oscillates. The asymptotic behavior of $R(\tilde{z})$
at $\tilde{z} \gg 1$ is as follows
\begin{equation}
\label{RAS}
R(\tilde{z}) \approx  \frac{\Gamma(2\kappa+2)}{\tilde{z}^{2\kappa+1}}
\sin(\tilde{z}-\pi \kappa) \; .
\end{equation} 
The spatial period of the oscillations of $R$ is 
$\delta x=\pi s \hbar/e V$ and it decreases with increasing voltage. 
It is to be emphasized that such a behavior of the correlator 
of $\delta G$ (slow decay and oscillations) is entirely due to 
interactions. In the absence of the interactions, the correlator 
(\ref{TCF}) would simply reproduce the correlator (\ref{DC})
of the fluctuations of $n(x)$, i.e. it would decay monotonously
at distance $\sim R_c$ which is much smaller than $\delta x$.

At $\tilde{z}=0$, the formula (\ref{NUCOR}) defines the variance of 
$\delta \nu$. It is convenient to  rewrite it for the relative 
magnitude of the  fluctuations of $\delta G$
\begin{equation}
\label{TCV}
\frac{(\overline{(\delta G)^2})^{1/2}}{\overline{G}}=
\frac{\pi^{3/4}}{4(2 \kappa+1)^{1/2}} \frac{s}{v_F}
\left(1-\frac{v_F^2}{s^2} \right)^2 
\frac{\left(\overline{\delta n^2}\right)^{1/2}}{n_0}
\left(\frac{\omega R_c}{s}\right)^{1/2}
\end{equation} 
Naturally, the variance $(\overline{\delta G^2)}^{1/2}$ is 
proportional to the magnitude  of the fluctuations of the density.
Less trivial is that (\ref{TCV}) contains two additional factors:
$(1-v_F^2/s^2)^2$ which is proportional to $V_0^2$ at small
$V_0$, and a small factor $(\omega R_c/s)^{1/2}$ (note that  (\ref{TCV})
is written in the limit $\omega \ll s/R_c$).
This factor determines the voltage dependence:
$(\overline{(\delta G)^2})^{1/2}/G \propto \sqrt{V}$.
Qualitatively, this dependence can be interpreted as follows.
Formula (\ref{FdDOS}) shows that all plasmon modes with frequencies 
smaller than $\omega$ contribute to $\delta \nu(x_0,\omega)$. 
Obviously, the major contribution comes from plasmons for which 
the center of localization lies within the localization radius 
from the point $x_0$. Let us consider some frequency  strip 
$(\omega_1- \Delta,\omega_1+\Delta)$ centered at 
$\omega_1 < \omega$ and with the width $\Delta \ll \omega_1$.
Then, the localization radius for all modes within this strip 
is approximately $l_{\omega_1}$. For a spatial interval of length 
$l_{\omega}$, the typical  frequency spacing between modes is
$1/ \overline{\rho} l_{\omega_1} \sim 1/ \tau_{\omega_1}$. 
Correspondingly, the average number of modes within the strip is 
$\Delta \tau_{\omega_1}$ and, hence, the relative fluctuation of this 
number is $ \sim (\Delta \tau_{\omega_1})^{-1/2}$.
The final estimate for the fluctuation of $\delta \nu $ emergies 
if we set $\Delta \sim \omega_1 \sim \omega$. Then, one gets 
$\delta \nu/\nu =\delta G / G \sim (\omega \tau_{\omega})^{-1/2}
\approx (k l_{\omega})^{-1/2}$. Substituting (\ref{MFG}) for 
$k l_{\omega}$, we reproduce the voltage dependence in (\ref{TCV}).
However, the estimate and the result of the calculation still differ 
by a factor $(1-v_F^2/s^2)$. This factor originates from the specific
details of the strucure of the eigenfunctions (see the comment after 
Eq. (\ref{Int_e})), and we cannot interpret it qualitatively. 
Note in conclusion of the section, that  since all frequencies from 
$0$ to $\omega$ give, roughly, comparable contributions to 
$\delta \nu(x_0,\omega)$, it will change significantly only when 
the frequency doubles. This means that  at fixed $x_0$  the characteristic 
period of fluctuations in $\delta G$ as a function of voltage is 
of the order of $V$.

\section{Conclusions}
\label{Con}
The main result of the present paper is that the randomness in the 
concentration of electrons in the Luttinger liquid causes a random 
component in the tunnel conductance which changes semiperiodically 
along the liquid, with the period depending on the applied bias. 
This behavior results from  the fact that at a fixed bias, V, 
the frequencies of plasmons, responsible for the correction to 
the density of states, are strictly limited by $eV/\hbar$.
Correspondingly, the oscillating behavior of the correlator 
of the fluctuations reflects the distribution  of density in a 
plasmon with maximal frequency.
If the conductance is studied  as a function of bias, the disorder 
would cause  fluctuations with  characteristic period 
of the order of $V$. The fluctuations disappear as the wavelength 
of the plasmon with frequency $eV/\hbar$ becomes smaller 
than the spatial scale  of the change of the concentration, $R_c$.
There is also a limitation from low voltages, imposed by the finite 
length of the liquid, $L$.
Namely, our calculation applies when the mean free path of a plasmon,
$l_{\omega}$, is smaller than $L$. At low enough frequencies
$l_{\omega}$ exceeds $L$ and the disorder does not play any role.
For such frequencies  the oscillations of the tunnel conductance 
with voltage have their origin in the size quantization of plasmons,
the period of oscillations being $\pi \hbar s/e L$. 
These oscillations were studied in Ref. \onlinecite{Na}.

Note finally, that the approach developed in the present paper can
be extended to the case of the multichannel Luttinger liquid 
\cite{Gla}. In the latter case, the scattering of plasmons 
between the channels, caused by a disorder, should be taken into account.

\acknowledgments

One of the authors (M.R.) is  grateful to I. L. Aleiner and A. I. Larkin
for a very useful discussion.

\begin{figure}
\label{fig1}
\caption{The correlation function of tunnel conductance is plotted
as a function of the dimensionless distance $\tilde{z}$ for $s/v_F=1.2$}
\end{figure}


\begin{references}
\bibitem{Note} Actually there is no qualitative understanding by now, 
why $F_1$ turns to zero at coinciding points \cite{A.P}.
\end{references}
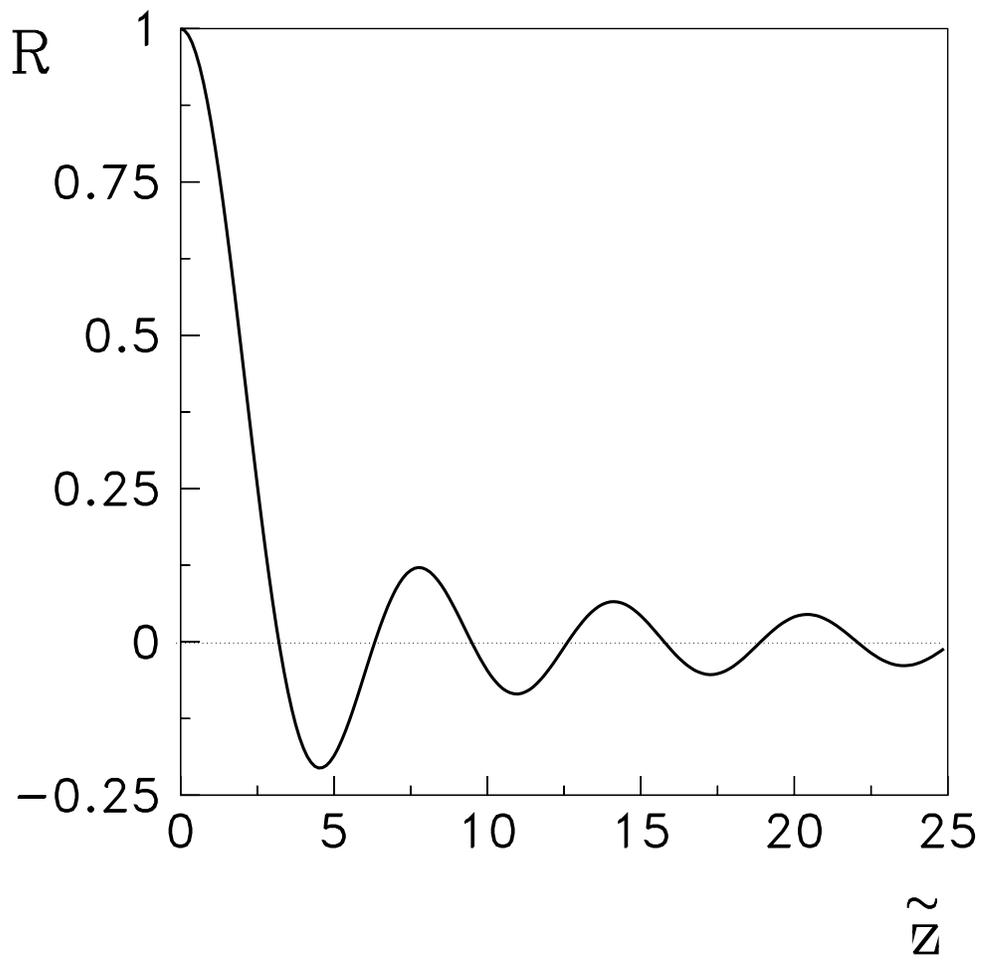
\end{document}